# $Yb^{3+}$ Speciation and Energy-Transfer Dynamics in Quantum-Cutting $Yb^{3+}$-Doped $CsPbCl_3$ Perovskite Nanocrystals and Single Crystals

Joo Yeon D. Roh,[1] Matthew D. Smith,[1] Matthew J. Crane,[1] Daniel Biner,[2] Tyler J. Milstein,[1] Karl W. Krämer,[2] Daniel R. Gamelin[1]*

[1]*Department of Chemistry, University of Washington, Seattle, WA 98195-1700, USA*

[2]*Department of Chemistry and Biochemistry, University of Bern, Freiestrasse 3, CH-3012 Bern, Switzerland*

*Email: gamelin@chem.washington.edu

**Abstract.** $Yb^{3+}$-doped inorganic metal-halide perovskites ($Yb^{3+}$:$CsPbX_3$, X = Cl, Br) have recently been discovered to display highly efficient quantum cutting, in which the energy from individual blue or UV photons absorbed by the material is re-emitted in the form of *pairs* of near-infrared photons by $Yb^{3+}$ dopants. Experimental photoluminescence quantum yields approaching 200% have been reported. As the first quantum-cutting materials that combine such high photoluminescence quantum yields with strong, broadband absorption in the visible, these materials offer unique opportunities for enhancing the efficiencies of solar technologies. Little is known about the fundamental origins of this quantum cutting, however. Here, we describe variable-temperature and time-resolved photoluminescence studies of $Yb^{3+}$:$CsPbCl_3$ in two disparate forms - colloidal nanocrystals and macroscopic single crystals. Both forms show very similar spectroscopic properties, demonstrating that quantum cutting is an intrinsic property of the $Yb^{3+}$:$CsPbX_3$ composition itself. Diverse $Yb^{3+}$ speciation is observed in both forms by low-temperature photoluminescence spectroscopy, but remarkably, quantum cutting is dominated by the same specific $Yb^{3+}$ species in both cases. Time-resolved photoluminescence measurements provide direct evidence of the previously hypothesized intermediate state in the quantum-cutting mechanism. This intermediate state mediates relaxation from the photogenerated excited state of the perovskite to the emissive excited state of $Yb^{3+}$, and hence is of critical mechanistic importance. At room temperature, this intermediate state is populated within a few picoseconds and has a decay time of only ~7 ns in both nanocrystalline and single-crystal $Yb^{3+}$:$CsPbCl_3$. The mechanistic implications of these observations are discussed. These results provide valuable information about characteristics of this unique quantum cutter that will aid its optimization and application in solar technologies.

# I.     INTRODUCTION

For several decades, researchers have strived to develop materials and technologies that can improve the efficiencies of solar cells beyond the so-called Schockley-Queisser thermodynamic limit of ca. 32% for a single-junction photovoltaic [1,2]. One strategy that has generated a lot of interest involves splitting the energies of individual solar photons to generate multiple lower-energy electron-hole pairs in the device, thus increasing the solar photocurrent and reducing thermalization losses. For example, certain organic chromophores such as tetracene have the capacity to split the energy of a high-energy photogenerated singlet excited state by populating two triplet excited states in neighboring molecules ("singlet fission"), each with roughly half of the initial photon's energy [3-5]. Similarly, semiconductor quantum dots (QD) can show "multiple exciton generation", in which a photogenerated upper excited state of the QD can cross-relax by promoting one (or more) additional electron(s) across the gap to yield two (or more) electron-hole pairs within the same QD, *i.e.*, bi- or multi-excitons [5-7]. In principle, these materials can then be used to generate photocurrents from high-energy solar photons at a greater rate than in traditional photovoltaics, but extraction of the excitation energy poses major challenges because of competing nonradiative recombination channels (*e.g.*, Auger), the need to introduce additional energy- or charge-transfer steps, or the need to re-engineer the underlying solar cell to integrate the new component at the expense of its baseline performance.

In phosphor research, an analogous photomultiplication process has been investigated in lanthanide-containing materials, referred to as "quantum cutting" [8-15]. A common rendition involves $Pr^{3+}$ and $Yb^{3+}$ co-doped into inorganic host crystals [9,11]. Photoexcitation of the $Pr^{3+}$ dopant with a blue photon is followed by non-radiative energy transfer to generate two excited $Yb^{3+}$ ions. A major advantage of quantum cutting is that the resulting $Yb^{3+}$ excited states are good photon emitters, in contrast with the dark triplet states generated *via* singlet fission. This advantage makes it feasible to extract the excitation energy photonically rather than *via* charge separation and collection or *via* a separate triplet-harvesting intermediate species. The two excited $Yb^{3+}$ ions generated by quantum cutting can simply re-emit the energy of the blue photon in the form of two NIR photons, both suitable for capture by an underlying photovoltaic. In these systems, however, $Pr^{3+}$ and other lanthanides are poor sensitizers because of their sharp and weak absorption features. Broadband sensitization has been explored but introduces additional loss channels.

Recently, $Yb^{3+}$-doped $CsPbX_3$ ($Yb^{3+}$:$CsPbX_3$, X = Cl, Br) has emerged as an attractive material capable of highly efficient quantum cutting [15-25], achieving photoluminescence quantum yields (PLQYs) approaching 200% [16,17]. First prepared as colloidal nanocrystals (NCs) [15,16], efficient quantum cutting has now also been demonstrated in both solution-processed [17] and vapor-deposited [21] $CsPbX_3$ thin films. A key distinction between this material and all previous quantum-cutting materials is its strong and broadband absorption at the energies relevant for quantum cutting. This feature allows much greater harvesting of solar photons by $Yb^{3+}$:$CsPbX_3$ than by all-lanthanide quantum cutters. Moreover, the composition tunability of the $CsPbX_3$ absorber material allows the absorption threshold to be tailored to minimize thermalization losses, and quantum-cutting energy efficiencies over 90% have been demonstrated for converting absorbed blue photons into emitted NIR photons [19]. Based on these properties, detailed-balance calculations have predicted double-digit (relative) improvements in the maximum theoretical power-conversion efficiencies of various photovoltaics when interfaced with these unique materials, including of state-of-the-art Si heterojunction solar cells [22]. Experimental results interfacing quantum-cutting $Yb^{3+}$-doped $CsPbX_3$ NCs with polycrystalline Si and CIGS photovoltaics have already demonstrated power-conversion efficiency enhancements as large as 20% (relative) [15,23]. Given the unique optoelectronic characteristics of these quantum-cutting materials and their preparative flexibility, other technologies such as high-performance transparent luminescent solar concentrators [18,20], high-efficiency NIR light-emitting diodes [24], or telecommunications phosphors [25] based on $Yb^{3+}$-doped $CsPbX_3$ and related compositions also become viable.

Because $Yb^{3+}$-doped metal-halide perovskites are a newly discovered composition of matter, the fundamental origins of their exceptional photophysics remain largely unresolved. One reason for their high quantum-cutting efficiency appears to be the extremely rapid depopulation of the photogenerated $CsPbX_3$ exciton upon introduction of $Yb^{3+}$ dopants. Transient-absorption measurements [16,17] have shown exciton depopulation within just a few picoseconds associated with $Yb^{3+}$ doping, making quantum cutting competitive with other non-productive exciton trapping or recombination processes. Curiously, energy transfer to $Yb^{3+}$ in other quantum cutters generally requires much more time than this because of the highly shielded *f*-shell valence orbitals involved in the $Yb^{3+}$ *f-f* excitation, exacerbated by the high ionicity of common host lattices. Using $La^{3+}$ as a surrogate trivalent dopant, photoluminescence (PL) from a shallow

"dopant-induced defect state" was observed, and this state was hypothesized to play a critical role as an intermediate state in this material's quantum cutting [16]. Transient absorption showed very similar picosecond exciton depopulation upon doping $CsPbCl_3$ NCs with $La^{3+}$ as found with $Yb^{3+}$ [16], supporting that hypothesis. The dopant-induced defect was proposed to arise from the need for charge compensation when substituting $Yb^{3+}$ for $Pb^{2+}$. Several charge-compensating defects could conceivably form in this material, and a charge-neutral defect cluster involving $Yb^{3+}$-$V_{Pb}$-$Yb^{3+}$ was hypothesized as a plausible motif by analogy with the "McPherson pairs" found in lanthanide-doped $CsCdX_3$ and related lattices [26,27]. Computational work supports the proposal that such a charge-neutral defect cluster can help to steer energy toward $Yb^{3+}$ dopants [28]. Despite the circumstantial evidence, however, there is to date no *direct* experimental evidence of the involvement of an intermediate state in the quantum cutting displayed by $Yb^{3+}$:$CsPbX_3$. Additional fundamental studies are required to unravel the properties of this unique material.

Here, we report results from variable-temperature and time-resolved PL (TRPL) studies of $Yb^{3+}$:$CsPbX_3$ NCs, as well as parallel results obtained for a $Yb^{3+}$:$CsPbX_3$ single crystal (SC) of macroscopic dimensions grown by the Bridgman method. Remarkably, these two disparate forms of the same composition show nearly indistinguishable spectroscopic characteristics, including nearly identical $Yb^{3+}$ *f-f* spectra that confirm that the same $Yb^{3+}$ species is responsible for quantum cutting in both nano- and macroscopic crystals. Most surprisingly, both materials also show nearly identical few-nanosecond rise times in the $Yb^{3+}$ PL generated by semiconductor photoexcitation. This result provides the first direct experimental evidence of a discrete intermediate state involved in the quantum cutting mechanism, and its preservation in both nano- and macroscopic crystals with very different surface-to-volume ratios and grown under very different conditions demonstrates its intrinsic origin. With this information in hand, the electronic-structure origins of quantum cutting in this material are discussed. The data further indicate the presence of upstream losses prior to energy capture by $Yb^{3+}$, as well as energy migration and trapping following quantum cutting, that both reduce the quantum-cutting efficiencies. The NCs show fewer losses in both steps, consistent with their higher PLQYs. These results improve our understanding of the photophysics and electronic structure of this unique material, and the insights from these measurements will help to inform future computational or experimental work including material optimization for device applications.

## II. EXPERIMENTAL

### A. Nanocrystal synthesis, single-crystal growth, and general materials characterization

$Yb^{3+}$:$CsPbCl_3$ NCs were synthesized as detailed previously [16]. Large crystals of $Yb^{3+}$-doped $CsPbCl_3$ were grown from melts of stoichiometric admixtures of precursors by the Bridgman technique. Additional synthesis and crystal-growth details are provided in the Supplemental Material [29]. Transmission electron microscopy (TEM) images were obtained using a FEI TECNAI F20 microscope operating at 200 kV. TEM samples were prepared by dropcasting NCs onto carbon-coated copper grids from TED Pella, Inc. Inductively coupled plasma-atomic emission spectroscopy (ICP-AES, PerkinElmer 8300) was used to determine elemental composition. NCs were digested in concentrated nitric acid overnight with sonication for ICP-AES. Powder X-ray diffraction (XRD) data were collected using a Bruker D8 Discover with a high-efficiency IμS microfocus X-ray source for Cu Kα radiation (50 kV, 1 mA). All $Yb^{3+}$ concentrations in $CsPbCl_3$ are reported as the B-site cation mole fraction (in percentage). For NC XRD, colloidal NCs were dropcast onto a silicon substrate. A portion of the Bridgman sample was powdered, and those microcrystals were used to obtain powder XRD data for this sample.

### B. Spectroscopic measurements

Samples for PL measurements were prepared by dropcasting colloidal NCs onto quartz discs, and by sealing a monolithic SC fragment inside a quartz tube under a partial pressure of helium gas. These samples were cooled to 5 K in a helium flow cryostat. Temperatures were varied between 5 and 295 K, and PL data were measured at each temperature using a 375 nm LED (0.3 μW/cm$^2$) for excitation. For NIR photoexcitation measurements, samples were excited with a CW Ti:Sapphire laser (Coherent Mira-HP in CW mode). PL spectra were measured using a LN$_2$-cooled silicon CCD camera mounted on a 0.3 m single monochromator with a spectral bandwidth of about 0.5 nm. Absolute PLQY measurements were performed at room temperature using a 5.3 inch teflon-based integrating sphere. The samples were directly excited with a 375 nm LED and attenuated with neutral density filters as needed, and the signal was measured using the CCD camera as described above. PLQY was calculated using Eq. (1), where $N$ indicates number of photons, $I$ indicates the spectrally corrected intensity of the emitted light, and $E$ indicates the spectrally corrected excitation intensity:

$$PLQY = \frac{N_{em}}{N_{abs}} = \frac{\int I_{sample}(\lambda) - I_{ref}(\lambda) d\lambda}{\int E_{ref}(\lambda) - E_{sample}(\lambda) d\lambda}. \quad (1)$$

The integrating sphere setup was routinely calibrated using well-characterized dye emission standards including coumarin 153, rhodamine 6G, and IR 140. All steady-state PL spectra were corrected for the instrument response.

Photoexcitation for TRPL measurements was provided by a liquid dye laser (Exalite 404 dye, $4.4 \times 10^{-4}$ M) pumped by an Ekspla Nd:YAG laser (355 nm) firing at a repetition rate of 50 Hz with a pulse width of about 30 ps. The fluence at the sample was held at ~100 nJ/cm$^2$ per pulse, which corresponds to about 0.01 absorbed photons per NC per pulse at room temperature. The NIR PL was focused into a monochromator with a spectral bandwidth of about 6 nm, detected by a Hamamatsu InGaAs/InP NIR PMT, and signals were recorded using a multichannel scaler or a digital oscilloscope.

## III. RESULTS AND ANALYSIS

### A. Structure

Figure 1 summarizes structural data collected for representative Yb$^{3+}$-doped CsPbCl$_3$ NCs ([Yb$^{3+}$] = 4.3%) and a Bridgman SC ([Yb$^{3+}$] = 2%, nominal). The TEM image in Fig. 1(a) shows nanocrystallites of Yb$^{3+}$-doped CsPbCl$_3$ with average edge lengths of ~11.6 nm (surveying ~350 NCs), consistent with previous results from the same synthesis method [16,19,20,30]. Figure 1(b) shows a photograph of a Yb$^{3+}$-doped CsPbCl$_3$ SC having ca. $3 \times 2 \times 2$ mm dimensions, sealed inside a quartz tube under a helium atmosphere. Powder XRD patterns collected from both the NCs and the SC sample (Fig. 1(c), powdered SC sample used) show intensities consistent with the perovskite crystal structure. CsPbCl$_3$ adopts the orthorhombic GdFeO$_3$-type structure at room temperature [31], undergoing a transition to the cubic phase at 37-47°C. The distortion of the room-temperature structure is small and gives rise to tiny additional reflections that are not visible on the scale of Fig. 1(c). As reported previously [16,17,19-21,30], Yb$^{3+}$-doping does not cause significant shifts of the XRD reflections relative to undoped CsPbCl$_3$ (Fig. 1(c)).

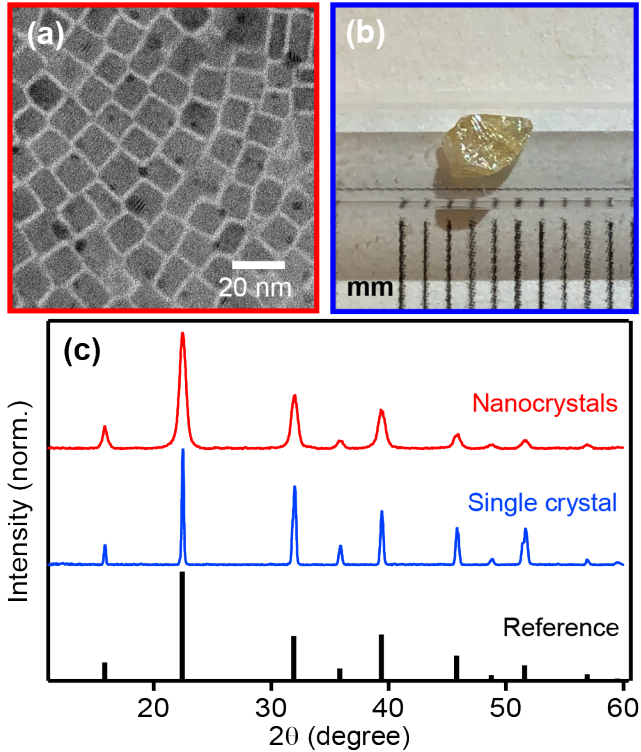

FIG. 1. (a) TEM image of 4.3% $Yb^{3+}$:$CsPbCl_3$ NCs and (b) a photograph of a 2% $Yb^{3+}$:$CsPbCl_3$ SC. Tick marks indicate mm spacings. (c) Powder XRD data collected for the same $Yb^{3+}$:$CsPbCl_3$ NCs (red) and SC (blue). Reference indices (black) are shown for the cubic high-temperature form of $CsPbCl_3$ (PDF 73-692), which is stable in bulk above ~37-47 °C [32]. Bulk $CsPbCl_3$ adopts the orthorhombic $GdFeO_3$-type structure (space group Pnma) at room temperature [31], giving rise to additional small reflections that are not visible on the scale of panel (c).

## B. Photoluminescence spectra

Figure 2 shows steady-state PL spectra of 1.7% $Yb^{3+}$:$CsPbCl_3$ NCs and the 2% $Yb^{3+}$:$CsPbCl_3$ SC of Fig. 1(b) measured at several temperatures from 5 to 295 K. The 5 K PL spectra of both samples (Fig. 2(a)) are highly structured, showing features characteristic of $Yb^{3+}$ $^2F_{5/2} \rightarrow {}^2F_{7/2}$ transitions in related chloride lattices. Notably, almost all of the peaks from the NC sample are directly correlated with peaks at the same energies and with the same relative intensities in the SC sample. The similarities between these spectra indicate similar environments for the perovskite-sensitized $Yb^{3+}$ ions in the two forms of the material. This observation is important because of the extremely large surface-to-volume ratios of the NCs compared to the bulk SC, and it demonstrates that the active $Yb^{3+}$ ions in the NCs are indeed bulk-like rather than, *e.g.*, bound to the NC surfaces.

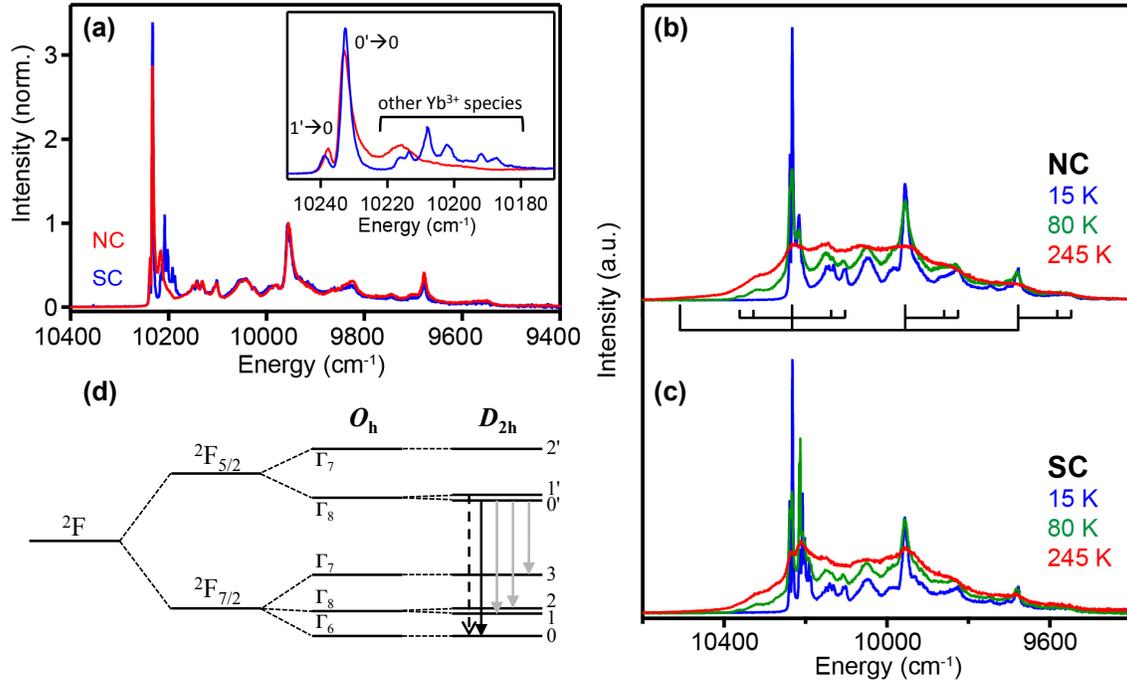

FIG. 2. (a) 5 K PL spectra of $Yb^{3+}$:$CsPbCl_3$ NCs (red, $[Yb^{3+}]$ = 1.7%) and the SC sample from Fig. 1 (blue, $[Yb^{3+}]$ = 2%), normalized to the peak at ~9960 $cm^{-1}$. Inset: The first set of peaks at highest energies in both samples at 5 K, showing the dominant 0' → 0 and 1' → 0 (hot-band) crystal-field components of the $Yb^{3+}$ $^2F_{5/2}$ → $^2F_{7/2}$ transition as well as a series of smaller peaks from other $Yb^{3+}$ species. Variable temperature PL spectra of (b) the NCs and (c) the SC from panel A measured at 15 K (blue), 80 K (green), and 245 K (red). $\lambda_{ex}$ = 375 nm for all panels. The spectrum in (b) includes markers denoting the periodic structure attributed to vibronic coupling, both on the Stokes and anti-Stokes (hot-band) sides of the first intense maximum. (d) Schematic illustration of $Yb^{3+}$ $f$-$f$ energy levels, including the low-symmetry splitting of $\Gamma_8$ components. The idealized reduced site symmetry is labeled as $D_{2h}$. The solid and dashed black arrows show the $^2F_{5/2}$ → $^2F_{7/2}$ 0' → 0 and 1' → 0 crystal-field transitions labeled in panel (a), respectively. The gray arrows indicate other low-temperature crystal-field origins that have not been conclusively identified in the spectrum.

Both PL spectra in Fig. 2(a) show their maximum peak intensities in a sharp feature at 10233 $cm^{-1}$, interpreted as an electronic origin within the crystal-field-split $Yb^{3+}$ $^2F_{5/2}$ → $^2F_{7/2}$ spectrum. In the cubic limit, the $^2F_{7/2}$ ground term is split by the crystal field into $\Gamma_6$, $\Gamma_8$, and $\Gamma_7$ levels, in order of increasing energy, and the $^2F_{5/2}$ excited term is split into $\Gamma_8$ and $\Gamma_7$ levels, such that this highest-energy origin at low temperature is associated with the $\Gamma_8(^2F_{5/2})$ → $\Gamma_6(^2F_{7/2})$ transition (*vide infra*). To lower energy, the next most prominent peak occurs at 9957 $cm^{-1}$, shifted ~276 $cm^{-1}$ from the first peak, and a third distinct peak is found another ~276 $cm^{-1}$ to lower energy with lower intensity. Each of these peaks appears to additionally show similar side bands at ~93 and 130 cm$^{-1}$. Such regularity in the spectral pattern suggests that this emission gains electric-dipole allowedness by the vibronic mechanism, despite high shielding of the *f* orbitals involved in these transitions [33,34], and hence that the quantum-cutting Yb$^{3+}$ ions reside at sites that lack a large odd-parity crystal-field component. In this mechanism, electric-dipole allowedness is enhanced by coupling the pure electronic transition with one quantum of an odd-parity local vibrational mode of the [YbCl$_6$]$^{3-}$ moiety. In support of this interpretation, we note that the energies here are very similar to those of the vibronic sidebands observed in the luminescence of Yb$^{3+}$-doped Cs$_2$NaHoCl$_6$ (86 cm$^{-1}$ ($\nu_6$), 108 cm$^{-1}$ ($\nu_4$), and 257 cm$^{-1}$ ($\nu_3$), all ungerade local modes) [35]. An additional broad luminescence peak is observed at ~10048 cm$^{-1}$ whose assignment is unclear. This feature occurs close in energy to the $\Gamma_8(^2F_{5/2}) \rightarrow \Gamma_8(^2F_{7/2})$ origin in cubic Yb$^{3+}$-doped Cs$_2$NaHoCl$_6$ [35] and could result from this electronic transition in somewhat reduced symmetry, for example due to a proximal charge-compensating defect. Alternatively, the $\Gamma_8(^2F_{5/2}) \rightarrow \Gamma_8(^2F_{7/2})$ origin may instead coincide with the intense peak at 9957 cm$^{-1}$ in Fig. 2(a), akin to the spectra of Yb$^{3+}$ in cubic fluoroperovskites [36]. Further studies on the SC sample and on halide-alloyed samples will be aimed at clarifying these assignments.

Both the NC and SC samples show an additional weak PL peak at 10239 cm$^{-1}$, *i.e.*, ~6 cm$^{-1}$ to *higher* energy of the 10233 cm$^{-1}$ $\Gamma_8(^2F_{5/2}) \rightarrow \Gamma_6(^2F_{7/2})$ origin. This feature is attributed to a thermal hot band reflecting a low-symmetry splitting of the emissive $\Gamma_8(^2F_{5/2})$ level into 0' and 1' crystal-field components. In support of this interpretation, we note that elevating the temperature from 5 to 15 K increases the intensity of this higher-energy band by nearly a factor of 3. Fitting this temperature dependence in relation to that of the 10233 cm$^{-1}$ intensity using a simple two-level Boltzmann model (see Supplemental Material [29]) yields an energy splitting that agrees well with the spectroscopic splitting and supports the peak assignment. Nearly identical results are obtained for the SC sample. This $\Gamma_8(^2F_{5/2})$ splitting energy is small compared to those of Yb$^{3+}$ ions in many other pseudo-octahedral crystalline sites (*e.g.*, ~45 cm$^{-1}$ in hexagonal CsMnCl$_3$ [37] and ~16 cm$^{-1}$ in hexagonal CsCdBr$_3$ [38,39]), from which it is inferred that the Yb$^{3+}$ ions involved in quantum cutting have a site symmetry that is close to octahedral.

In addition to populating the $\Gamma_8(^2F_{5/2})$ hot band, raising the sample temperature also broadens all of the individual PL features observed at 5 K and introduces additional hot bands at higher energies. These hot bands occur with the same energies identified in the low-temperature spectrum, supporting the assignment of these features as vibronic. For example, the 80 K spectrum of the NCs (Fig. 2(b)) shows clear hot bands ~93 and 130 cm$^{-1}$ to higher energy of the 10233 cm$^{-1}$ low-temperature origin, and the 245 K spectrum shows an additional broad shoulder that would be consistent with a ~276 cm$^{-1}$ hot band (see Supplemental Material [29]), although the spectral breadth at this temperature impedes a concrete assignment. Moreover, the Yb$^{3+}$ PL decay times decrease substantially with increasing temperature (*vide infra*) while the integrated PL intensities increase (Fig. 2(b, c)), consistent with a vibronic electric-dipole intensity-gaining mechanism and hence Yb$^{3+}$ centrosymmetry. Concomitantly, the excitonic PL intensity of the NCs decreases with increasing temperature (see Supplemental Material [29]), as we reported previously for other NC samples [16].

EPR measurements of Mn$^{2+}$-doped CsPbCl$_3$ SCs have indicated essentially orthorhombic B-site point symmetries of $D_2$, $C_{2v}$, or $D_{2h}$ at all temperatures below the cubic phase transition (~320 K) [40], of which only $D_{2h}$ has the inversion symmetry implied by the PL spectra and variable-temperature PL data described above. Similarly, EPR measurements of Gd$^{3+}$-doped CsPbCl$_3$ SCs have also suggested a centrosymmetric point group at room temperature, assigned as $C_{2h}$ [41]. In both cases, reduction from orthorhombic to monoclinic and possibly lower site symmetries are formally required in lower-temperature phases, but these distortions appear to be minor [40,41]. With these considerations and the spectroscopic observations described above, we assume an effective site symmetry of either $D_{2h}$ or $C_{2h}$ for Yb$^{3+}$ in CsPbCl$_3$ at all temperatures examined here. Fig. 2(d) summarizes the Yb$^{3+}$ crystal-field splittings so far deduced for this system, assuming an idealized site symmetry of $D_{2h}$.

The main difference between the NC and SC spectra is in the set of peaks within the first ~50 cm$^{-1}$ of the first intense maximum, *e.g.*, at ca. 10230 cm$^{-1}$ (Fig. 1(a), inset). Here, the NCs appear to show a simpler spectrum, with only one clear side peak at 10216 cm$^{-1}$, whereas the SC shows a multitude of maxima in this region. Site-selective photoexcitation measurements on the SC sample demonstrate that these various sharp features come from different Yb$^{3+}$ species, and at least 5 distinct species can be deduced (see Supplemental Material [29]). Because the charge of the Yb$^{3+}$ dopants is different from that of the B-site Pb$^{2+}$ cations, additional defect formation is always required to achieve charge neutrality, and the additional PL lines are thus attributed to Yb$^{3+}$ "trap" sites with different charge-compensation motifs. These Yb$^{3+}$ trap species are only resolved near the first electronic origin, where spectral broadening is smallest. In support of the interpretation of these peaks as due to other $Yb^{3+}$ species, we note that some of their intensities increase markedly with increasing temperature in the SC spectra but not in the NC spectra (Fig. 2). This observation is consistent with thermally assisted energy migration and capture by $Yb^{3+}$ traps in the SC sample, as also reflected in the increasingly multi-exponential $Yb^{3+}$ decay (*vide infra*) and in the low PLQY (~20%) of the SC sample. Although numerous, these $Yb^{3+}$ traps account for < ~30% of the integrated quantum-cutting PL at low temperature (see Supplemental Material [29]). The fact that the majority of quantum-cutting PL intensity arises from just the 10230 cm$^{-1}$ species even though that species is not dominant in any of the PL spectra obtained with direct $Yb^{3+}$ photoexcitation strongly suggests that quantum cutting occurs non-statistically, *i.e.*, the 10230 cm$^{-1}$ $Yb^{3+}$ species is disproportionately active in quantum cutting. Notably, the 10230 cm$^{-1}$ quantum-cutting species also shows by far the most intense vibronic structure among the species observed by site-selective excitation (see Supplemental Material [29]). Overall, we conclude that quantum cutting is dominated by the same specific $Yb^{3+}$ species in both NC and SC forms of this material.

### C. Photoluminescence dynamics

Figure 3 plots room-temperature NIR PL decay curves measured for the $Yb^{3+}$:CsPbCl$_3$ NCs ([$Yb^{3+}$] = 1.7%) and SC ([$Yb^{3+}$] = 2%) of Fig. 2. The NCs (Fig. 3(a)) show nearly monoexponential PL decay following CsPbCl$_3$ photoexcitation, with $\tau_{decay}$ ~ 1.5 ms. The inset to Fig. 3(a) reveals a distinct rise in the $Yb^{3+}$ PL following the excitation pulse, having a time constant of ~8 ns. Similar data are found for the $Yb^{3+}$:CsPbCl$_3$ SC (Fig. 3(b)), except that the $Yb^{3+}$ PL decay in this sample is multiexponential. Measurement at short times again reveals a ~7 ns rise time for $Yb^{3+}$ PL following CsPbCl$_3$ photoexcitation.

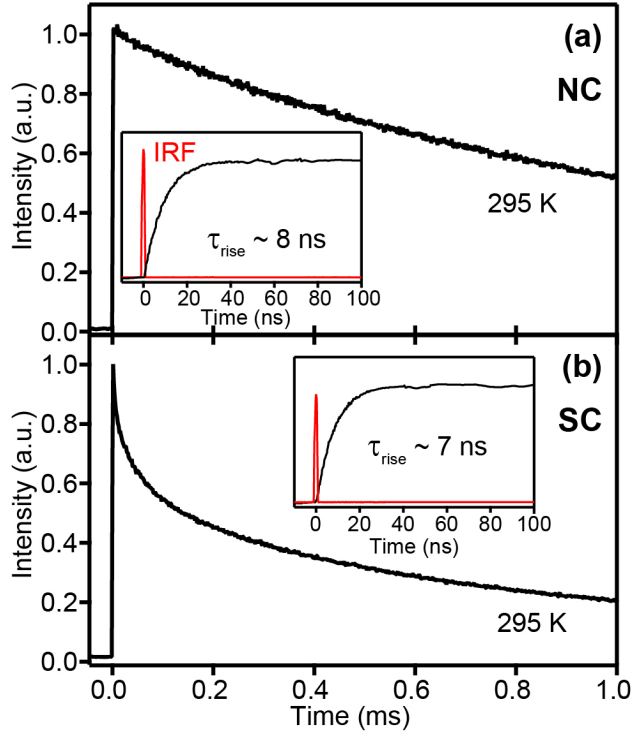

FIG. 3. Representative room-temperature TRPL data collected for (a) NCs ([Yb$^{3+}$] = 1.7%) and (b) a SC of Yb$^{3+}$:CsPbCl$_3$ ([Yb$^{3+}$] = 2%(nom.)), respectively). Inset: First 100 ns of the room-temperature TRPL trace, showing a distinct rise at short times. The red curve plots the experimental instrument response function (IRF). $\lambda_{ex}$ = 404 nm, $\lambda_{em}$ = 985 nm.

Figure 4 plots TRPL data measured for the Yb$^{3+}$:CsPbCl$_3$ NCs and SC from Fig. 2 at temperatures from 5 to 295 K following interband photoexcitation of the CsPbCl$_3$. Figure 4(a) plots the Yb$^{3+}$ PL intensities at short times on a linear intensity scale, and Fig. 4(b) plots the same PL decay over much longer times and on a logarithmic intensity scale. The NC PL decay curves in Fig. 4(b) are all nearly monoexponential and the decay rate accelerates by roughly a factor of 2 upon warming from 5 to 295 K, even though the integrated PL intensity increases over this temperature range. This temperature dependence is consistent with the conclusion drawn above that the Yb$^{3+}$ excited *via* quantum cutting occupies a centrosymmetric lattice site and that thermal energy accelerates the *radiative* Yb$^{3+}$ $^2F_{5/2}$ → $^2F_{7/2}$ transition through vibronic coupling. Figure 4(a) shows that the rise of the NC Yb$^{3+}$ PL gets much slower as the temperature is lowered. Whereas the PL rise at room temperature is too fast to be observed in the window of Fig. 4(a) (but see Fig. 3(a), inset, and the Supplemental Material [29]), the PL rise at 5 K is biexponential with its dominant (~80%) component having a time constant of ~40 ns (*vide infra*) followed by a much slower component (~20%) with a time constant of ~50 μs.

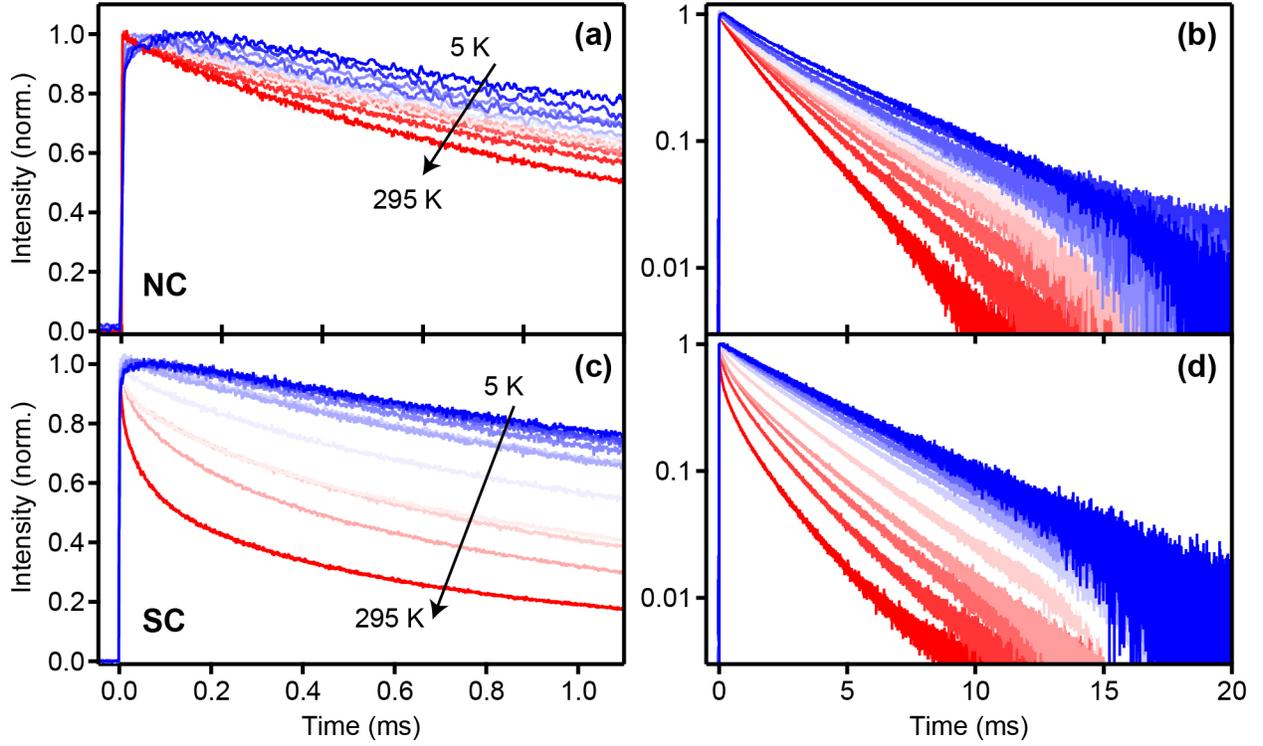

FIG. 4. Variable-temperature TRPL traces measured for (a, b) NCs and (c, d) a SC of $Yb^{3+}$:$CsPbCl_3$ (1.7% and 2%(nom.) $Yb^{3+}$, respectively), from 5 K (blue) to room-temperature (red). Temperatures: 5, 15, 30, 45, 60, 80, 100, 125, 150, 200, 250, 295 K. $\lambda_{ex}$ = 404 nm, $\lambda_{em}$ = 982 nm (at 5-150 K), 985 nm (at 200-295 K).

Figure 4(c, d) plots analogous data collected for the $Yb^{3+}$:$CsPbCl_3$ SC. Similar trends are observed, with slower $Yb^{3+}$ PL rise and decay dynamics at lower temperatures. The SC data are complicated by additional processes that make the $Yb^{3+}$ PL decay curves multiexponential, most evident in the appearance of a prominent PL decay component with a time constant of ~100 μs at room temperature that is absent at lower temperatures. Despite these specific differences, the SC TRPL data (Fig. 4(c, d)) are generally very similar to those measured for the NCs (Fig. 4(a, b)).

Figure 5 summarizes the data from Figs. 2 and 4 for both the $Yb^{3+}$:$CsPbCl_3$ NCs (Fig. 5(a-c)) and the $Yb^{3+}$:$CsPbCl_3$ SC (Fig. 5(d-f)). Figure 5(a) plots the temperature dependence of the integrated steady-state $Yb^{3+}$ PL intensity, along with the temperature dependence of the PL decay time, $\tau_{decay}$, obtained by fitting the data to a single-exponential function. Figure 5(d) plots analogous variable-temperature data for the $Yb^{3+}$:$CsPbCl_3$ SC. Here, the figure plots the amplitude-weighted average decay time, $\tau_{decay}(avg)$ (Eq. (2)), where $n$ indexes the decay components), obtained from biexponential fits of the PL decay curves (see Supplemental Material [29]);

$$\tau_{decay}(avg) = \frac{\sum A_n \tau_n^2}{\sum A_n \tau_n}. \qquad (2)$$

Figure 5(a) and 5(d) both show the same trend of increasing $Yb^{3+}$ PL intensity with increasing temperature, more than doubling over this temperature range. In parallel, the $Yb^{3+}$ PL decay time decreases from over ~4 ms at 5 K to ~1.5 ms at 295 K for both samples. These trends are similar to the NC results reported previously [16]. The PL temperature dependence described by Fig. 5(a, d) also suggests that PL saturation, which is directly linked to the long $Yb^{3+}$ excited-state lifetime [30], may be partially alleviated in solar cells, which operate at about 80 °C (~350 K).

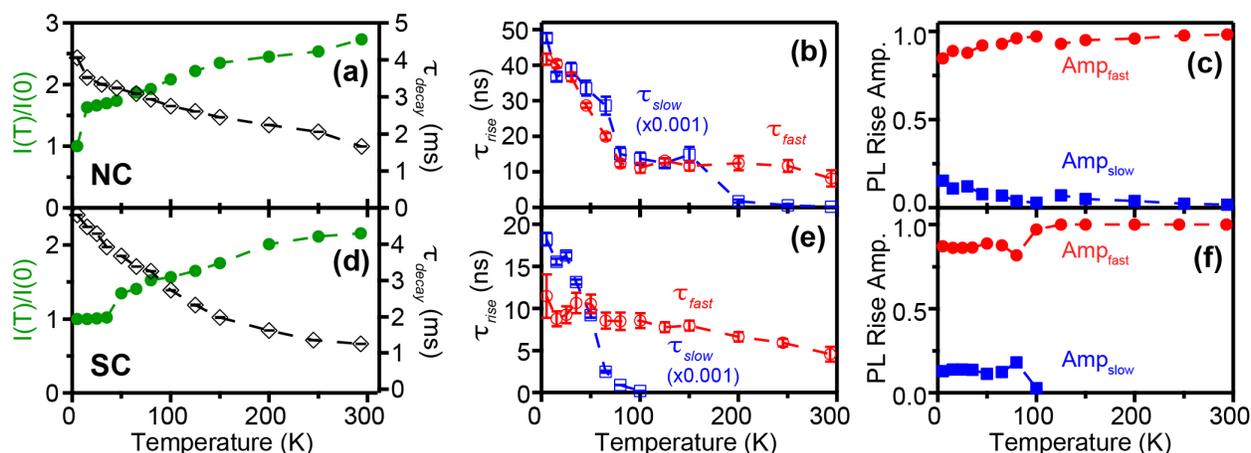

FIG. 5. Scatter plots summarizing the variable-temperature PL data in Figs. 2 and 4. (a, d) Integrated NIR PL intensities (green) and decay times (black) for the $Yb^{3+}$:CsPbCl$_3$ NCs (a) and SC (d) plotted vs temperature. For the SC, $\tau_{decay}$ reflects the average PL decay time obtained from biexponential fitting, as described by Eq. (2). (b, c, e, f) Fast (red) and slow (blue) NIR PL rise times and amplitudes measured for the $Yb^{3+}$:CsPbCl$_3$ NCs (b, c) and SC (e, f), plotted vs temperature.

**TABLE I. Key results from analysis of $Yb^{3+}$ photoluminescence rise and decay dynamics at 5 K and 295 K following pulsed photoexcitation of the CsPbCl$_3$ host in ~2% $Yb^{3+}$:CsPbCl$_3$ nanocrystal and single-crystal samples (Figures 4 and 5).**

|  | Nanocrystals | | Single Crystal | |
|---|---|---|---|---|
|  | 5 K | 295 K | 5 K | 295 K |
| Decay: $\tau_{decay}$ | 4.1 ms | 1.7 ms | 4.4 ms | 1.3 ms |
| Rise: $\tau_{fast}$ (% amplitude) | 41.6 ns (81%) | 7.9 ns (98%) | 11.8 ns (87%) | 7.0 ns (100%) |
| Rise: $\tau_{slow}$ (% amplitude) | 47.7 µs (19%) | 100 ns (2%) | 18.2 µs (13%) | -- |

The bi-exponential $Yb^{3+}$ PL rise data for the $Yb^{3+}$:CsPbCl$_3$ NCs (Fig. 4(a)) were fitted and the time constants and amplitudes of each component are plotted separately in Fig. 5(b, c). At 5 K, ~80% of the NC PL rise occurs with a time constant of $\tau_{fast}$ ~ 42 ns, and the remaining 20% has a rise time of $\tau_{slow}$ ~ 48 μs. Both rise components accelerate (Fig. 5(b)) and the fast component dominates more (Fig. 5(c)) with increasing temperature. The amplitude of the slow rise decreases from ~20% at 5 K to close to 0% at 100 K, and only a fast rise component of $\tau_{fast}$ ~ 8 ns is discernable at 295 K. Very similar results are again found for the SC (Fig. 5(e, f)). Here, at 5 K, the fast rise component has a fractional amplitude of ~87% and a time constant of only ~12 ns, whereas the slow component (~13%) is characterized by $\tau_{slow}$ ~ 18 μs. By room temperature, the slow rise is no longer discernable and the fast rise has accelerated to $\tau_{fast}$ ~ 7 ns (Fig. 3(b), inset). Energy transfer to $Yb^{3+}$ following semiconductor photoexcitation is thus very fast at room temperature in both the NC and SC $Yb^{3+}$:CsPbCl$_3$ samples.

## IV. DISCUSSION

### A. Comparison of NCs and SC

A primary observation from the results and analysis presented above is that the NCs and SC are essentially identical to one another, despite their massively different surface-to-volume ratios and the very different reaction conditions under which these two forms of $Yb^{3+}$:CsPbCl$_3$ were prepared (solution precipitation *vs* Bridgman growth). Structurally, the powder XRD data for these two forms are basically indistinguishable. Even more diagnostic are the high-resolution low-temperature PL data; both samples show remarkably similar sensitized $Yb^{3+}$ PL peak energies and intensities, indicating that the $Yb^{3+}$ species active in quantum cutting in both forms are the same. Surprisingly, even the dynamics of energy transfer to $Yb^{3+}$ following CsPbCl$_3$ photoexcitation are essentially identical in the NCs and SC. Both forms show an $Yb^{3+}$ PL rise time constant of about 7 ns at room temperature. These observations highlight the conclusion that quantum cutting in the $Yb^{3+}$-doped CsPbCl$_3$ composition does not require any other properties of NCs or the other granular forms that have been explored previously [17,21], such as surfaces, grain-boundary defects, or spatial exciton confinement.

Despite these overwhelming similarities, some differences are still observed between the NCs and SC, and these differences can be informative. In high-resolution PL spectra, for example, the two samples show different distributions of $Yb^{3+}$ traps. These traps reflect a variety of charge-compensation motifs generated upon substitutional doping of $Yb^{3+}$ in both the NCs and the SC, but with different distributions of these traps between the two forms. Such a difference is not surprising given the very different synthesis conditions used for preparing the two materials. The SC sample shows evidence of thermally assisted energy migration and capture by these traps, which is manifested in their different PL temperature dependence, the temperature dependence of the $Yb^{3+}$ quantum-cutting PL decay dynamics, and ultimately the relatively low PLQY of the SC sample (~20%). Additionally, the NCs show excitonic PL at low temperatures that disappears when the temperature is raised, but the SC sample shows no excitonic PL at any temperature (see Supplemental Material [29]). This result suggests that nonproductive exciton recombination is more competitive in the SC sample than in the NCs, and it is also consistent with the low PLQY of the SC sample (~20%). Notably, the similar $Yb^{3+}$ PL rise times for these two samples suggest that these losses occur *prior* to energy capture by the intermediate defect state. Despite these specific differences between $Yb^{3+}$-doped $CsPbCl_3$ NCs and SC, the primary photophysical characteristics of these two forms of the material are almost identical.

### B. Quantum-cutting energy-transfer dynamics

Previously, we demonstrated that $Yb^{3+}$ doping of $CsPbCl_3$ NCs and solution-deposited granular thin films [16,17] introduces a new few-picosecond exciton depopulation channel attributed to quantum cutting. An intermediate state associated with a hypothesized dopant-induced defect was proposed to mediate energy transfer from the perovskite to $Yb^{3+}$ [16]. The observation here of a ~7 ns rise at room temperature in the $Yb^{3+}$ PL following pulsed interband photoexcitation, and the fact that this rise is slower than the exciton depopulation time, provides the first direct evidence that energy transfer from the perovskite to $Yb^{3+}$ indeed proceeds *via* such an intermediate state, supporting the general energy-transfer mechanism proposed previously [16]. Although the PL data presented above also reveal other $Yb^{3+}$ species in both NC and SC samples, presumably associated with different charge-compensation motifs, the data point to just one specific species as dominant in quantum cutting.

A slower (μs) pathway for $Yb^{3+}$ sensitization is also observed in the PL rise dynamics at low temperatures, but the PL is almost entirely dominated by the fast (ns) pathway in both the NCs and the SC, and so we focus our discussion on this process. An interesting fundamental question pertains to the electronic-structure origins of the fast energy-transfer rate. In most other quantum-cutting systems, the actual quantum-cutting energy-transfer step involves weak multipolar coupling between formally electric-dipole-forbidden *f—f* transitions of the energy donor (*e.g.*, $Pr^{3+}$ [9,11], $Tb^{3+}$ [8], $Er^{3+}$ [12], $Tm^{3+}$ [10], $Nd^{3+}$ [13], $Ho^{3+}$ [14]) and the quantum-cutting acceptor (typically $Yb^{3+}$). Consequently, quantum-cutting time constants ranging from ~6 μs – 4 ms are typically observed. More similar to the present system are the energy-transfer times observed in molecularly sensitized (non-quantum-cutting) $Ln^{3+}$ compounds, which typically show energy transfer from ligand singlet states to bound $Ln^{3+}$ ions on the timescale of 100s of ps to several ns [42]. This rapid energy transfer involves a Dexter-type exchange mechanism instead of the multipolar coupling mechanism active in typical lanthanide pairs. The absence of near-band-edge emission at room temperature (see Supplemental Material [29]) is an indication that the oscillator strength of the intermediate state is too small for its radiative recombination to compete with energy transfer to $Yb^{3+}$, consistent with a Dexter-type energy transfer mechanism. We thus propose that energy transfer in $Yb^{3+}$-doped $CsPbCl_3$ proceeds *via* an exchange-mediated mechanism that involves coupling between the dopant-induced defect and two $Yb^{3+}$ ions. Because Dexter-type energy transfer is a short-range effect, the dopant-induced defect must be close to both $Yb^{3+}$ acceptor ions, and it may in fact be this defect that differentiates the PL spectrum of the $Yb^{3+}$ sites involved in quantum cutting from the spectra of the other $Yb^{3+}$ sites in the crystal (Fig. 2a). Overall, these findings are generally consistent with the proposal [16] of a $Yb^{3+}$-$V_{Pb}$-$Yb^{3+}$ charge-neutral defect cluster at the heart of quantum cutting, although the precise microscopic defect structure remains unknown.

Figure 6 summarizes the energy flow upon perovskite photoexcitation. Efficient quantum cutting involves two consecutive energy-transfer steps: the first from the exciton to a shallow dopant-induced defect ($k_{defect}$), and the second from that defect state to a pair of $Yb^{3+}$ ions ($k_{QC}$). The dopant-induced defect state depopulates the excitonic state on a single-picosecond timescale and can compete with all other radiative and nonradiative processes affecting the exciton population [16,17]. The quantum-cutting step ($k_{QC}$) is also very rapid, with $1/k_{QC} \sim 7$ ns at room temperature. Both the NCs and SC show increasing integrated $Yb^{3+}$ PL intensities upon increasing the temperature from 5 to 295 K, despite more non-radiative decay of the $Yb^{3+}$ excited state ($k_{decay}$) at elevated temperatures in the SC. This observation indicates that the branching ratios for one or both of the two upstream quantum-cutting energy-transfer steps ($k_{\text{defect}}$ and $k_{\text{QC}}$) must become more favorable with increasing temperature. This conclusion is supported by the observation of decreasing excitonic PL intensity with increasing temperature in the NCs.

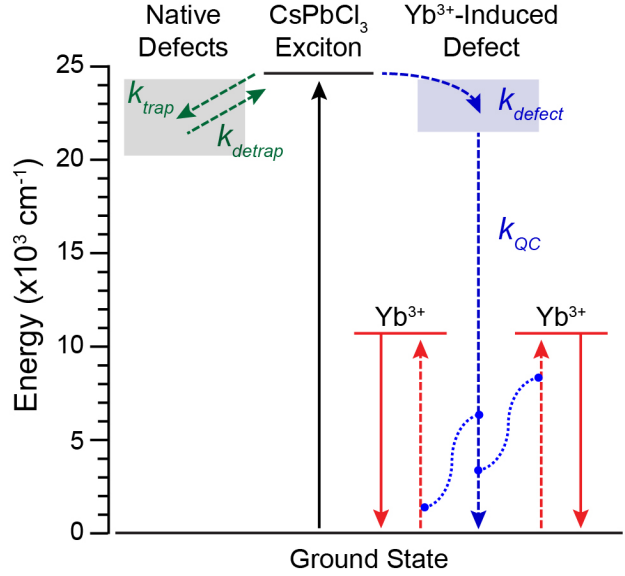

FIG. 6. Quantum-cutting energy transfer in $Yb^{3+}$-doped $CsPbCl_3$. Photoexcitation of $CsPbCl_3$ (black up arrow) is followed by intense $Yb^{3+}$ $f$–$f$ luminescence (red down arrows) after quantum cutting. The majority of the quantum cutting proceeds following the blue arrows, *via* energy capture by a dopant-induced defect ($k_{\text{defect}}$) and then simultaneous energy transfer to two $Yb^{3+}$ dopants ($k_{\text{QC}}$). This process takes < 10 ns at room temperature and < 50 ns at liquid-helium temperatures. A minority slow pathway is also observed at low temperatures that is proposed to involve temporary trapping and detrapping by native defects (green arrows). This process takes ~20 to 50 μs at liquid-helium temperatures. Solid arrows indicate radiative processes and dashed arrows indicate nonradiative processes.

Such an energy-transfer process can be considered in the framework of Fermi's golden rule, Eq. (3), which describes $k_{\text{QC}}$ as a function of the donor/acceptor electronic coupling ($M_{\text{DA}}$) and spectral overlap ($\rho$):

$$k_{QC} = \frac{2\pi}{\hbar} |M_{DA}|^2 \rho. \tag{3}$$

In $Yb^{3+}$-doped $CsPbCl_3$, $k_{\text{QC}} = 1/\tau_{\text{rise}}$ (the fast rise component), the donor in the quantum-cutting step is the dopant-induced defect state, and the acceptor is a pair of $Yb^{3+}$ ions that are both coupled to this defect state, and hence in proximity to the defect. For simplicity, we consider this $Yb^{3+}$ pair as formally a single "dimeric" acceptor with states at $2 \times E_{f\text{-}f}$, although a three-center formalism may be more appropriate for quantitative description.

We have previously explored the role of $\rho$ in Yb$^{3+}$:CsPb(Cl$_{1-x}$Br$_x$)$_3$ quantum cutting experimentally, using anion exchange to tune the donor energy [19]. Because the dopant-induced defect state is shallow, $E_{\text{donor}} \sim E_g$. For $E_g > 2 \times E_{f-f}$, the NIR PLQY is roughly independent of $E_g$. Given the narrowness of the $f$—$f$ absorption, $\rho$ must be small at large values of $E_g$ and quantum cutting must be phonon assisted, but $k_{\text{QC}}$ must nonetheless remain sufficiently large to maintain efficient quantum cutting. A steep drop in the NIR PLQY is observed when $E_g < \sim 2 \times E_{f-f}$, attributed to reduction of $\rho$ to zero, *i.e.*, a quantum-cutting threshold energy [17,19].

The electronic-coupling term ($M_{\text{DA}}$) in this system is of high fundamental interest, but it is as yet poorly understood. From the above discussion of energy-transfer rates in other systems, quantum cutting in Yb$^{3+}$:CsPbX$_3$ proceeds *via* a Dexter-type exchange mechanism. In this scenario, the electronic coupling between the partially localized dopant-induced defect state and Yb$^{3+}$ acceptors is likely mediated by Yb$^{3+}$-X$^-$ covalency. Computational work [28] has suggested that a proximal Pb$^{2+}$ ion of a right-angle Yb$^{3+}$-V$_{\text{Pb}}$-Yb$^{3+}$ defect cluster possesses increased electron density that could play an important role in facilitating this quantum-cutting step. In the construct of second-order perturbation theory, Yb$^{3+}$-X$^-$ covalency reflects configuration interaction between the Yb$^{3+}$(4$f$) states and low-lying halide-to-Yb$^{3+}$ charge-transfer (LMCT) excited states, which formally involve photoinduced electron transfer from the valence band to Yb$^{3+}$. Other recent experimental work has led to the proposal that this quantum cutting involves electron trapping by Yb$^{3+}$ to form a discrete Yb$^{2+}$ intermediate [25]. The position of the Yb$^{3+/2+}$ level relative to the perovskite band edges is not known. The NIR PL spectra in Fig. 2 are consistent with Yb$^{3+}$ coordinated by 6 Cl$^-$ anions in a pseudo-octahedral [YbCl$_6$]$^{3-}$ cluster, and the first LMCT excited states of comparable [YbCl$_6$]$^{3-}$ centers are not observed until $\sim$36,000 cm$^{-1}$ (compared with $E_g \sim$ 26,000 cm$^{-1}$ for CsPbCl$_3$) [43,44]. These LMCT transitions shift down to $\sim$29,000 cm$^{-1}$ in [YbBr$_6$]$^{3-}$ (compared with $E_g \sim$ 19,000 cm$^{-1}$ for CsPbBr$_3$) [45,46], but remain >1 eV higher in energy than the perovskite energy gap. This consideration, in conjunction with the experimental observation of new near-band-edge defect emission upon doping CsPbCl$_3$ with redox-inactive La$^{3+}$, points to a more general dopant-induced defect state rather than Yb$^{2+}$ as the quantum-cutting intermediate. Further combined experimental and theoretical advances will be required to fully understand the electronic structure of this unique material.

Finally, we address the origins of the slow component observed in the Yb$^{3+}$ PL rise dynamics. The observation of a biexponential PL rise demonstrates the existence of two separate pathways for energy transfer from the perovskite to $Yb^{3+}$. Although the slow rise only accounts for $<\sim 20\%$ of the $Yb^{3+}$ PL at 5 K and is not observable at room temperature, it suggests the presence of a competing process that intercepts and temporarily stores the photoexcitation energy in a separate metastable state before eventually transferring it to $Yb^{3+}$. Metastable charge-separated states have previously been postulated for undoped $CsPbX_3$ (X = $Cl^-$, $Br^-$) NCs and thin films, formed by hole trapping on the timescale of tens of picoseconds [47-49], and substantial delayed luminescence involving a metastable charge-separated state has also been reported for both SC and NC forms of the related hybrid perovskite, $CH_3NH_3PbBr_3$ [50,51]. We hypothesize that the slow quantum-cutting process observed here reflects rapid but reversible trapping in a similar metastable charge-separated state, as also illustrated in Fig. 6 ($k_{trap}$ and $k_{detrap}$). Thermal detrapping from this metastable state accelerates this slow quantum-cutting pathway at elevated temperatures, and by room temperature it is no longer distinguishable. We note that the precise microscopic steps in this slow pathway remain unclear; the pathway may involve equilibrium between excitonic and metastable excited states alone, as in delayed excitonic luminescence and depicted in Fig. 6, or it could conceivably involve energy transfer to $Yb^{3+}$ directly from the metastable state.

## V. CONCLUSION

Despite their very different synthetic origins and surface-to-volume ratios, $Yb^{3+}$:$CsPbCl_3$ NC and SC samples show remarkably similar spectral and photophysical characteristics. The structural and spectroscopic data reported here are consistent with substitutional $Yb^{3+}$-doping at the lattice B-site of perovskite $CsPbCl_3$. This aliovalent substitution requires additional defect formation to maintain charge neutrality, and several distinct $Yb^{3+}$ species are observed spectroscopically. Quantum cutting in both crystal forms is largely dominated by the same specific $Yb^{3+}$ species, however. This species appears to be close to centrosymmetric with nearly octahedral site symmetry and it appears to display relatively strong vibronic coupling in its *f-f* luminescence.

TRPL measurements have provided the first direct evidence of an intermediate state in the energy transfer from $CsPbCl_3$ to $Yb^{3+}$. In both NC and SC samples, a fast rise time of about 7 ns is observed in the room-temperature $Yb^{3+}$ PL following perovskite photoexcitation. This observation solidifies the prior hypothesis [16] that quantum cutting involves consecutive steps of (*i*) energy capture by a dopant-induced defect, followed by (*ii*) energy transfer from this defect to two $Yb^{3+}$ ions. The first step depopulates the photogenerated perovskite excited state within a few picoseconds, localizing the excitation in proximity to the $Yb^{3+}$ dopants, and the second step captures the excitation energy at $Yb^{3+}$ within a few nanoseconds. The nanosecond $Yb^{3+}$ PL rise time suggests that energy transfer from the intermediate dopant-induced defect state to $Yb^{3+}$ is mediated by exchange. Both energy-transfer steps accelerate with increasing temperature, becoming more competitive with other non-productive trapping or recombination processes. The rapid energy localization and capture described here are responsible for the high efficiency of quantum cutting in this material, making it kinetically competitive with all other processes following photoexcitation. An additional slower rise component (µs) of the $Yb^{3+}$ PL is observed at low temperatures, attributed to a second sensitization pathway that involves temporary storage of excitation energy in a native metastable trap state, similar to the scenario responsible for delayed excitonic luminescence. The results presented here provide new insights into the fundamental spectroscopic and photophysical properties of $Yb^{3+}:CsPbX_3$ quantum cutters that will help the development of this material for future solar and photonic applications.

**Acknowledgments**

This research was supported by the National Science Foundation (NSF) through DMR-1807394 and through the UW Molecular Engineering Materials Center, a Materials Research Science and Engineering Center (DMR-1719797). This work was also supported by the State of Washington through the UW Clean Energy Institute (to JDR), and to the Washington Research Foundation (to MJC). Part of this work was conducted at the UW Molecular Analysis Facility, a National Nanotechnology Coordinated Infrastructure site supported in part by the NSF (ECC-1542101), the University of Washington, the Molecular Engineering and Sciences Institute, the Clean Energy Institute, and the National Institutes of Health. The authors thank Werner Kaminsky for assistance with single-crystal X-ray crystallography, Daniel Kroupa for measuring the PLQY of the single-crystal sample, and Kyle Kluherz for TEM imaging.

*Supplementary Material for*

# Yb[3+] Speciation and Energy-Transfer Dynamics in Quantum-Cutting Yb[3+]-Doped CsPbCl$_3$ Perovskite Nanocrystals and Single Crystals

Joo Yeon D. Roh,[1] Matthew D. Smith,[1] Matthew J. Crane,[1] Daniel Biner,[2] Tyler J. Milstein,[1] Karl W. Krämer,[2] Daniel R. Gamelin[1]*

[1]*Department of Chemistry, University of Washington, Seattle, WA 98195-1700, USA*
[2]*Department of Chemistry and Biochemistry, University of Bern, Freiestrasse 3, 3012-Bern, Switzerland*
*Email: gamelin@chem.washington.edu

**Methods**

    **Materials.** For nanocrystal (NC) syntheses, n-hexane (99%, Sigma Aldrich), 1-octadecene (ODE, 90%, Sigma Aldrich), oleic acid (OA, 90%, Sigma Aldrich), oleylamine (OAm, 70%, Sigma Aldrich), lead acetate trihydrate (Pb(OAc)$_2$·3H$_2$O, 99.999%, Sigma Aldrich), cesium acetate (CsOAc, 99.9%, Sigma Aldrich), trimethylsilyl chloride (TMS-Cl, 98%, Acros Organics), and ethyl acetate (EtOAc, 99%, Sigma Aldrich) were used as received unless otherwise noted. As-received ytterbium acetate hydrate (Yb(OAc)$_3$·xH$_2$O, 99.9%, Alfa Aesar) was refluxed in glacial acetic acid for 1 hr and stored in a desiccator prior to use. For single crystal (SC) growth, cesium chloride (CsCl, 99.995%, Merck KGaA) was dried in vacuum at 200°C. Lead chloride (PbCl$_2$) was prepared from lead carbonate (PbCO$_3$, 99.999%, Alfa) and hydrochloric acid (HCl, 30%, suprapure, Merck KGaA). PbCl$_2$ was sublimed in vacuum at 450°C. Ytterbium chloride (YbCl$_3$) was synthesized from ytterbium oxide (Yb$_2$O$_3$, 99.9999%, Metall Rare Earth Ltd), sublimed ammonium chloride (NH$_4$Cl, 99.8%, Merck), and HCl acid [1]. YbCl$_3$ was sublimed in vacuum at 800°C. All starting materials and products were handled under strictly dry and oxygen-free conditions in a glove box (MBraun, Garching) or closed apparatus.

    **Nanocrystal synthesis.** NCs of 1.7% Yb[3+]:CsPbCl$_3$ were synthesized as detailed previously [2]. In a typical synthesis, Yb(OAc)$_3$ (12 mg, 0.034 mmol) and Pb(OAc)$_2$·3H$_2$O (76 mg, 0.200 mmol) were combined in a round bottom flask containing ODE (5 mL), OA (1 mL), OAm (0.25 mL), and 1 M CsOAc in EtOH (0.28 mL) and degassed on a Schlenk line at 110 °C for 1 hr. The solid precursors dissolved in this solution at ca. 110 °C. The reaction vessel was then flushed with N$_2$ and heated to 240 °C, whereupon a solution of 0.2 mL TMS-Cl in 0.5 mL ODE was quickly injected resulting in rapid NC formation. To quench the reaction, the vessel was immediately cooled using a room-temperature water bath. The crude NC solution was then centrifuged and n-hexane was added to the pellet to resuspend the NCs. The NCs were flocculated out of solution with ethyl acetate and centrifuged again for 5 min. The NCs were once again resuspended in n-hexane and centrifuged for 20 min. The supernatant of this solution was collected and contained the final nanocrystal product. Colloidal CsPbCl$_3$ NCs are not hygroscopic like their bulk counterparts, and they can be handled and stored in air without degradation over timescales relevant to the experiments reported here. These NC samples were stored in the dark in ambient atmosphere.

    **Single-crystal growth.** Large crystals of Yb[3+]-doped CsPbCl$_3$ were grown from melts of stoichiometric admixtures of CsCl, PbCl$_2$, and YbCl$_3$ by the Bridgman technique. The starting materials were sealed in a silica ampoule under vacuum and molten up at 650°C. Then, the

temperature was set to 625°C and crystals were grown by slow cooling over about 10 days. $Yb^{3+}$ concentrations are reported as the nominal concentrations of the initial powder mixture. Because the starting material $YbCl_3$ is hygroscopic, all sample handling and storage was done under inert atmosphere. Bulk $CsPbCl_3$ is stable in dry air. For a good thermal contact and in order to avoid any potential deterioration, crystals were sealed under inert helium atmosphere in silica ampules for spectroscopic measurements.

**Absorption.** Absorption spectra were measured using an Agilent Cary 5000 spectrometer operating in transmission mode.

**NIR photoluminescence excitation (PLE).** PLE spectra were measured at room temperature using an Edinburgh FLS 1000 photoluminescence spectrometer equipped with a 450 W xenon lamp as an excitation source and a spectral bandwidth of 1 nm. NIR PL was focused into a monochromator with a spectral bandwidth of 1 nm, detected by a Hamamatsu InGaAs/InP NIR photomultiplier.

**Additional Data and Analysis**

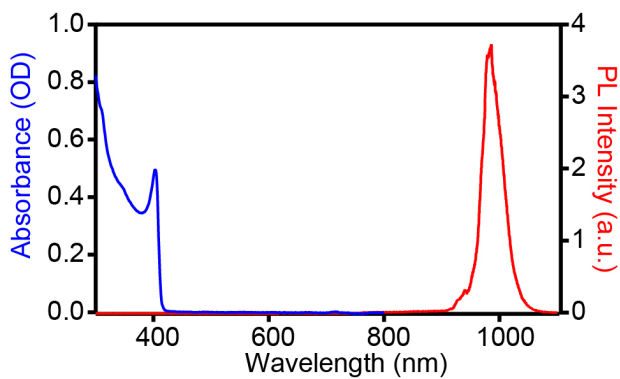

FIG S1. Overview room-temperature absorption (blue) and photoluminescence (PL) (red) spectra of 1.7 % $Yb^{3+}$:$CsPbCl_3$ NCs. $\lambda_{ex}$ = 375 nm.

S-2

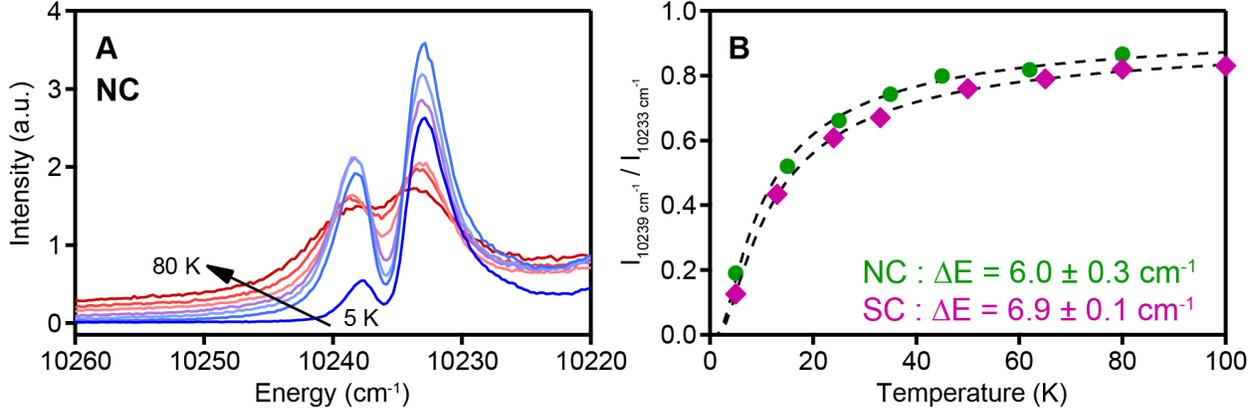

FIG S2. (A) Variable-temperature PL spectra of $Yb^{3+}$:$CsPbCl_3$ NCs from Figs. 2 and 4 of the main text, measured from 5 (blue) to 80 K (red). (B) The ratio of PL intensities of the hot band (10239 cm$^{-1}$) and the first electronic origin (10233 cm$^{-1}$) measured for the NCs (green) and the SC (purple), plotted as a function of temperature and fitted (dashed black) using the Boltzmann distribution equation of Eq. (S1). $\lambda_{ex}$ = 375 nm.

The intensity ratios plotted in Fig. S2B *vs* temperature (*T*) were fitted using Eq. (S1), where $I_i$ and $I_0$ denote peak intensities at 10239 and 10233 cm$^{-1}$, respectively, and *ΔE* is the floating variable. The fits yield *ΔE* = 6.0 ± 0.3 and 6.9 ± 0.1 cm$^{-1}$ for the NC and SC samples, respectively. These fit results agree well with the $\Gamma_8(^2F_{5/2})$ splitting energy of ~6 cm$^{-1}$ measured spectroscopically (10239 - 10233 cm$^{-1}$).

$$\frac{I_i}{I_0} \propto e^{\frac{\Delta E}{k_b T}} \tag{S1}$$

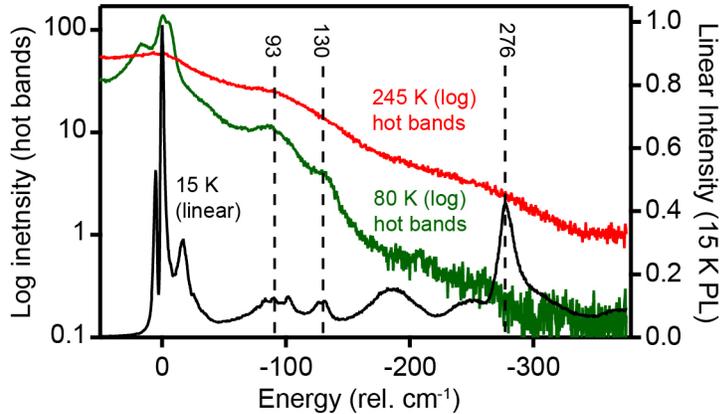

FIG S3. The PL spectra of the 1.7% $Yb^{3+}$:$CsPbCl_3$ NCs from Fig. 2B of the main text, replotted to show direct comparison of the 15 K quantum-cutting PL spectrum with the PL hot bands observed at 80 and 245 K. The 15 K PL spectrum is plotted on a linear y scale and the hot-band spectra are plotted on a logarithmic *y* scale. The vertical lines are guides to the eye to indicate coincidences between cold and hot band energies. $\lambda_{ex}$ = 375 nm.

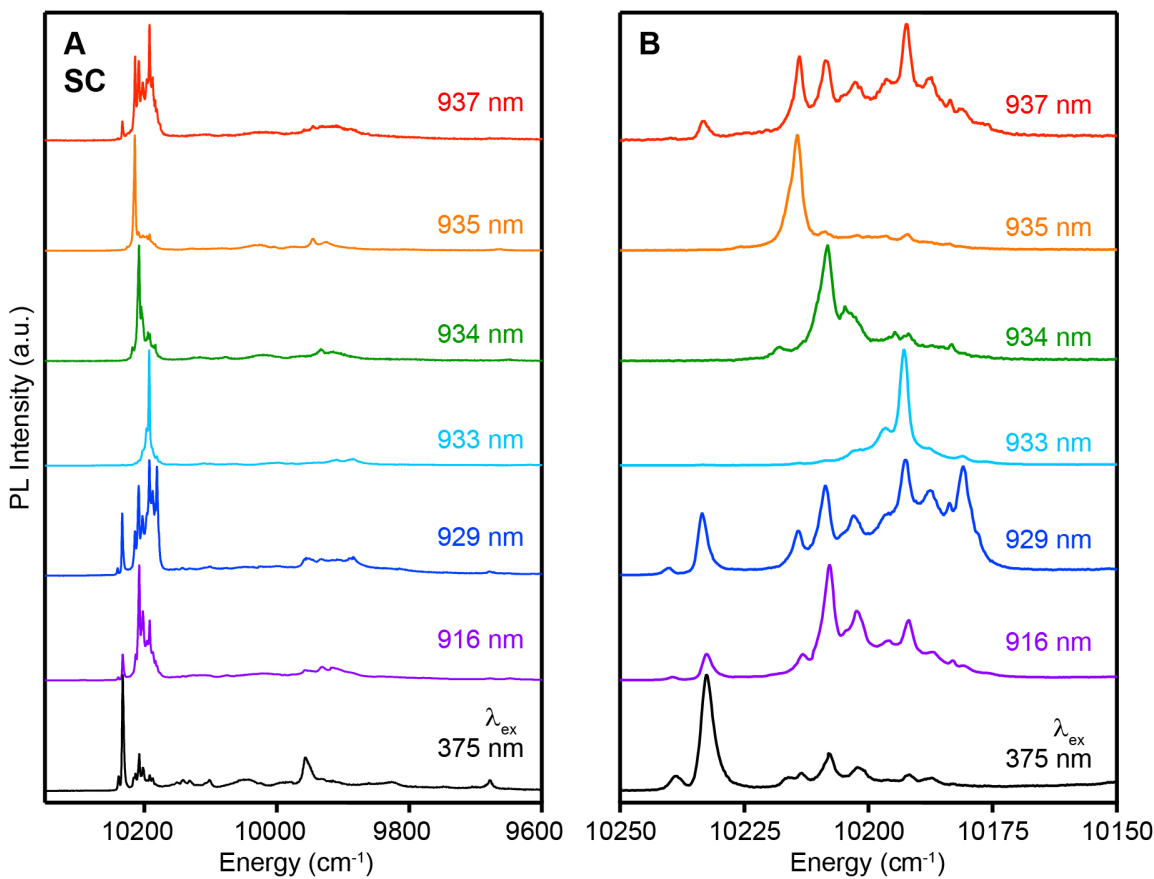

FIG S4. (A) 5 K PL spectra of the $Yb^{3+}$:$CsPbCl_3$ SC measured using site-selective laser excitation at 916 (purple), 929 (dark blue), 933 (blue), 934 (green), 935 (orange), and 937 (red) nm. The perovskite-sensitized quantum-cutting $Yb^{3+}$ PL spectrum excited at 375 nm (black) is included for comparison. (B) An expanded view of the set of peaks at highest energies in panel (A).

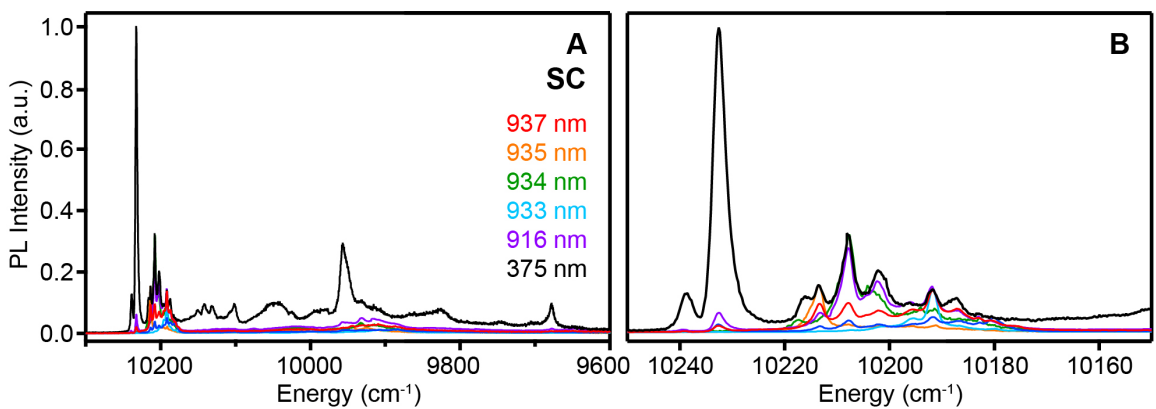

FIG S5. (A) 5 K PL spectra of the $Yb^{3+}$:$CsPbCl_3$ SC from Fig. S4 (colored), scaled to reflect the maximum possible contribution of each to the quantum-cutting PL spectrum (black). (B) An expanded view of the set of peaks at highest energies from panel (A).

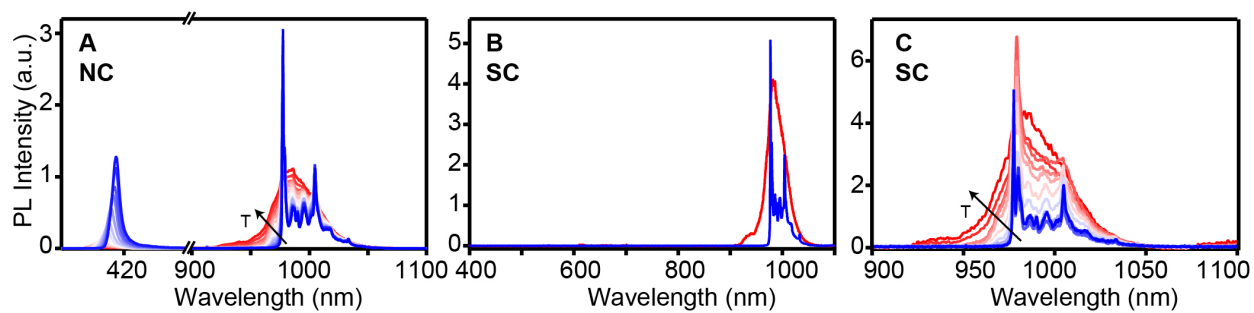

FIG S6. Variable-temperature PL spectra of (A) $Yb^{3+}$:$CsPbCl_3$ NCs ($[Yb^{3+}]$ = 1.7%) and (B, C) the $Yb^{3+}$:$CsPbCl_3$ SC ($[Yb^{3+}]$ = 2%(nom.)) measured from 5 (blue) to 295 K (red). There was no excitonic PL from the $Yb^{3+}$:$CsPbCl_3$ SC detectable at any temperature. $\lambda_{ex}$ = 375 nm.

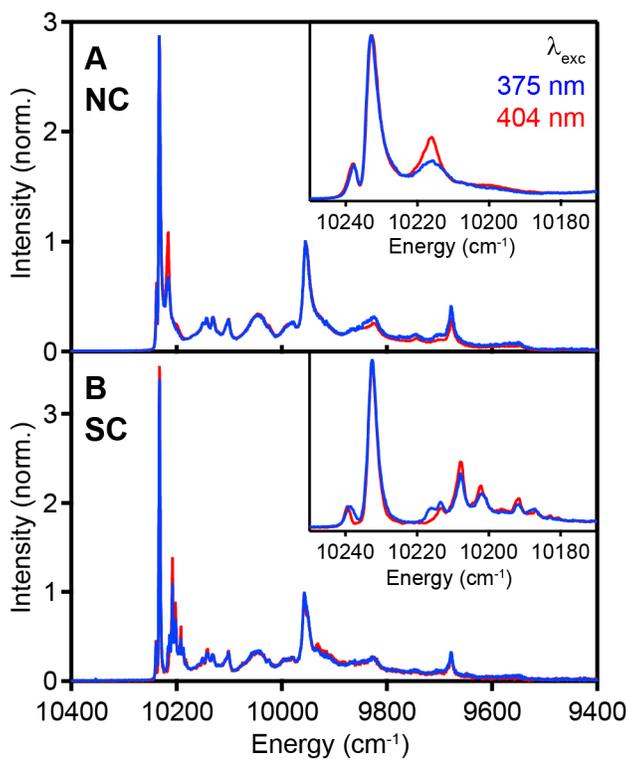

FIG S7. 5 K PL spectra of (A) $Yb^{3+}$:$CsPbCl_3$ NCs and (B) the $Yb^{3+}$:$CsPbCl_3$ SC, normalized to the peak at ~9960 cm$^{-1}$, measured using excitation at 375 (blue) and 404 nm (red). Insets: Expanded views of the set of peaks at highest energies. The spectra collected at the two different excitation wavelengths are very similar but not identical.

Average Yb$^{3+}$ PL decay times for the Yb$^{3+}$:CsPbCl$_3$ SC were determined by fitting the PL decay traces in Fig. 4B of the main text to a biexponential function. The amplitude-weighted average of decay time constants, $\tau_{decay}$(avg), are given by Eq. (S2), where $A_n$ denotes the amplitude of the $n$th decay component, $\tau_n$. These weighted averages are plotted in Fig. 5D of the main text. The complete fitting results for the Yb$^{3+}$:CsPbCl$_3$ SC are summarized in Table S1.

$$\tau_{decay}(avg) = \frac{\sum A_n \tau_n^2}{\sum A_n \tau_n} \qquad (S2)$$

**TABLE S1. Fitting parameters obtained from analysis of Yb$^{3+}$ PL decay data for the 2% Yb$^{3+}$:CsPbCl$_3$ SC.**

| Temp. (K) | $A_1$ | $\tau_1$ (ms) | $A_2$ | $\tau_2$ (ms) | $\tau_{decay}$(avg) (ms) |
|---|---|---|---|---|---|
| 5 | 0.527 | 2.98 | 0.473 | 5.70 | 4.70 |
| 15 | 0.523 | 2.74 | 0.478 | 5.44 | 4.48 |
| 25 | 0.405 | 2.39 | 0.595 | 4.91 | 4.29 |
| 30 | 0.181 | 1.46 | 0.820 | 4.12 | 3.93 |
| 45 | 0.177 | 1.24 | 0.823 | 3.84 | 3.67 |
| 60 | 0.164 | 0.85 | 0.836 | 3.50 | 3.38 |
| 80 | 0.137 | 0.72 | 0.863 | 3.34 | 3.25 |
| 100 | 0.206 | 0.48 | 0.795 | 2.83 | 2.73 |
| 125 | 0.331 | 0.38 | 0.669 | 2.47 | 2.32 |
| 150 | 0.345 | 0.45 | 0.655 | 2.15 | 1.98 |
| 200 | 0.391 | 0.26 | 0.609 | 1.75 | 1.62 |
| 250 | 0.447 | 0.20 | 0.553 | 1.48 | 1.36 |
| 295 | 0.476 | 0.13 | 0.524 | 1.35 | 1.25 |

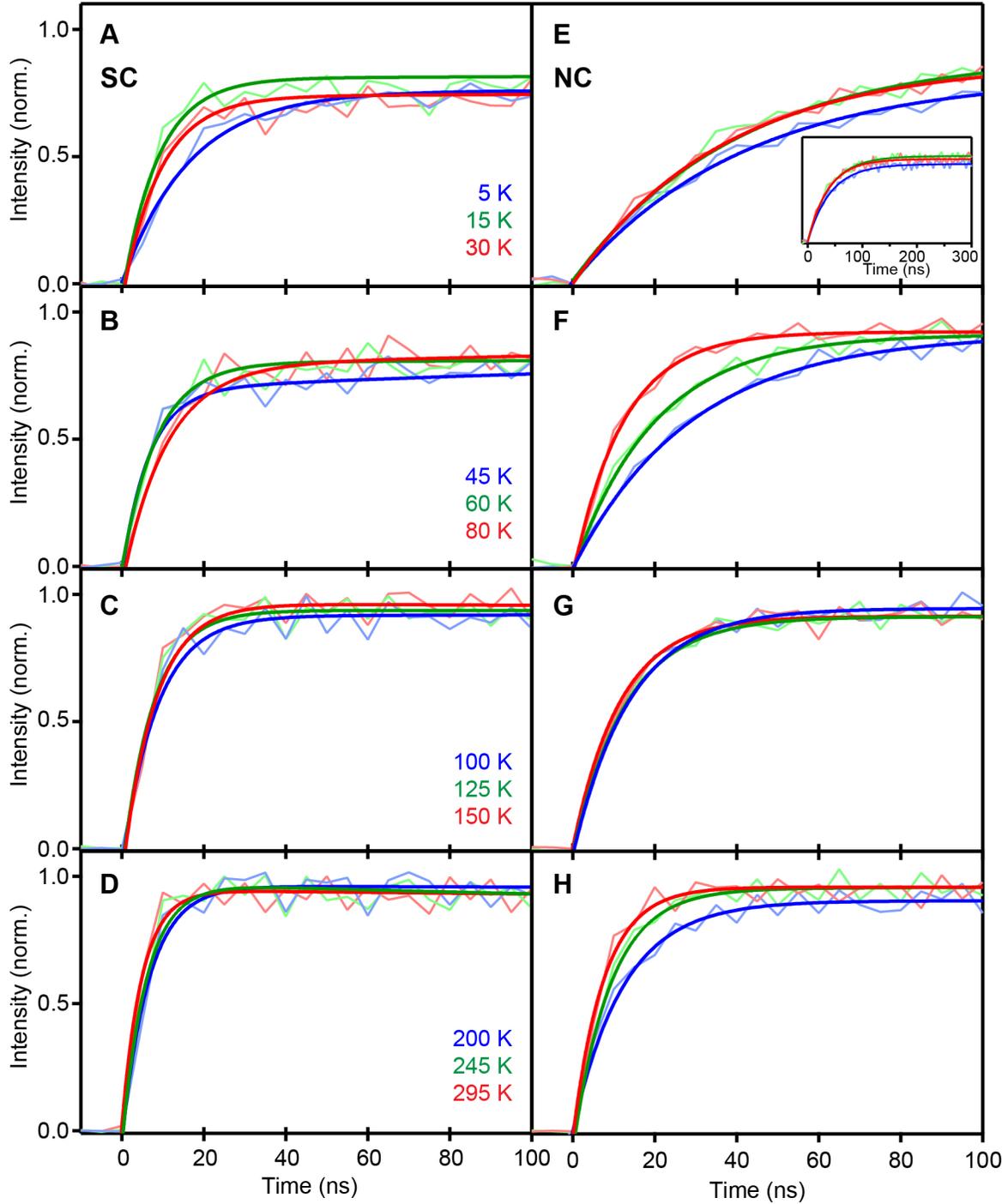

FIG S8. First 100 ns of TRPL traces (faint lines) measured at various temperatures from 5 to 295 K, and bi-exponential fits (solid lines) of each experimental trace. (A, B, C, D) Data for the 2.0%(nom.) $Yb^{3+}$:$CsPbCl_3$ SC, and (E, F, G, H) data for the 1.7% $Yb^{3+}$:$CsPbCl_3$ NCs. Inset to panel (E): First 300 ns of the TRPL traces and fits for the NC data measured at 5, 15, and 30 K, showing the complete rise. $\lambda_{ex}$ = 404 nm, $\lambda_{em}$ = 982 nm (at 5-150 K), 985 nm (at 200-295 K).

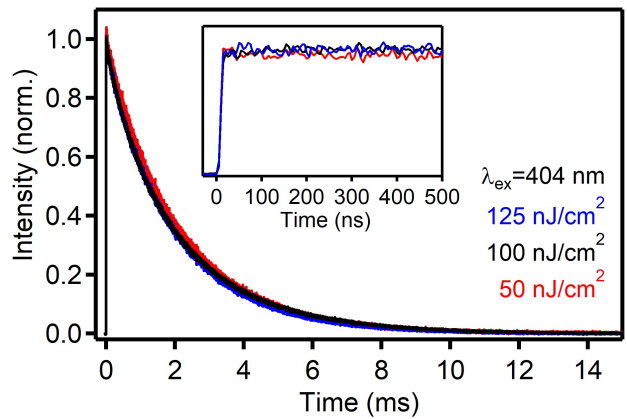

FIG S9. Excitation-fluence dependence of $Yb^{3+}$ PL decay measured for the 1.7% $Yb^{3+}$:$CsPbCl_3$ NCs at room temperature. $\lambda_{ex}$ = 404 nm, $\lambda_{em}$ = 985 nm. Inset: Expanded view of the first 500 ns of each $Yb^{3+}$ PL decay curve shown in the main figure.

**Yb$^{3+}$ NIR Photoluminescence Excitation (PLE) Spectra.** Numerous publications have reported PLE measurements of both doped and undoped perovskite NCs that appear to show decreasing PLQYs at short excitation wavelengths [3-7]. The large extinction coefficients of CsPbX$_3$ make PLE measurements susceptible to artifacts both experimental in origin and stemming from incorrect data representation. Experimental artifacts come from the fact that the absorption length in concentrated samples is short and changes substantially when the wavelength is scanned above the perovskite absorption edge, resulting in inconsistent PL collection efficiencies under normal measurement conditions. To illustrate, Figure S10 plots PLE spectra collected for the same Yb$^{3+}$:CsPbCl$_3$ NCs at two different concentrations, and shows that at low concentration the first PLE maximum aligns well with the first absorption maximum but at high concentration the first PLE maximum appears red-shifted and the PLE does not continue to rise at shorter wavelengths as the absorption spectrum does. These two observations are both artifacts and can be eliminated by using an integrating sphere during the PLE measurements (see below).

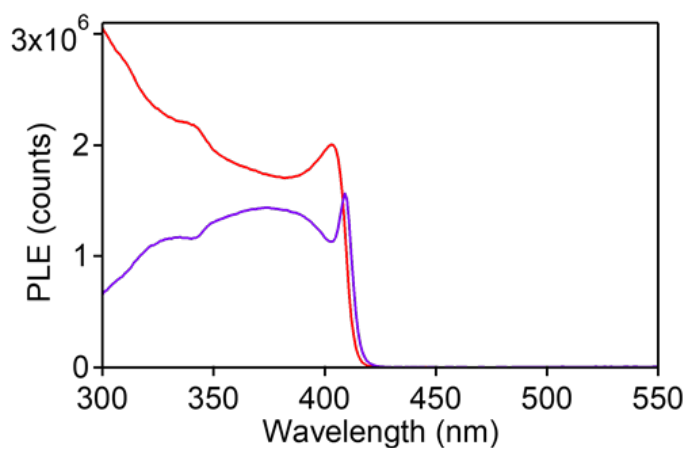

FIG S10. PLE spectra of low (red) and high (purple) concentration colloidal 0.3% Yb$^{3+}$:CsPbCl$_3$ NC samples suspended in hexane, measured without using an integrating sphere. The optical densities of these two samples were 0.22 and 1.88 at the first absorption maximum (403 nm), using a 1.0 cm pathlength cuvette.

Furthermore, PLE intensities reflect sample absorptance, not absorbance, where absorptance ($\alpha$) is defined as $\alpha = 1 - R - T$, $R$ is reflectance, and $T$ is transmittance. The absorptance approaches unity as the sample's effective concentration or path length increases, suppressing spectral structure at high optical density. To illustrate, Figure S11 plots Yb$^{3+}$ NIR PLE spectra of the same Yb$^{3+}$:CsPbCl$_3$ NCs collected at a series of different NC concentrations, and compares these with the absorptance spectra of the same samples. The PLE and absorptance plots agree well in every case. Only in the limit of large $T$ do the shapes of the absorptance and absorption spectra resemble one another closely, and in this limit the PLE also resembles the absorption spectrum.

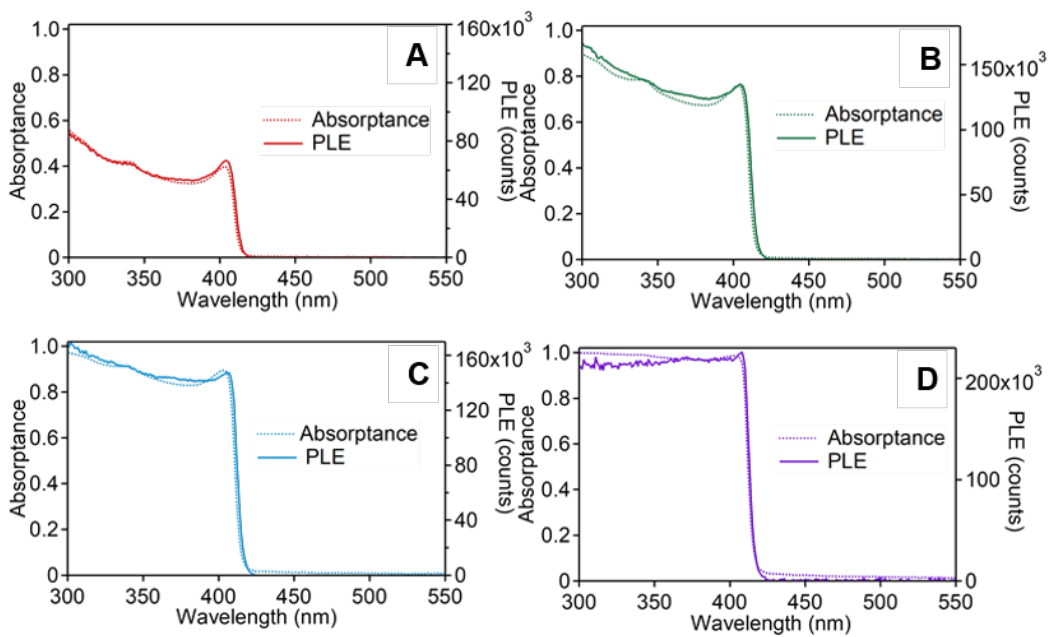

FIG S11. Absorptance and PLE spectra of colloidal 0.3% $Yb^{3+}$:$CsPbCl_3$ NCs at different NC concentrations in hexane. The optical densities at the first absorption maximum (403 nm) were 0.22, 0.62, 0.98, and 1.88 for samples A-D, respectively, using a 1.0 cm pathlength cuvette. PLE spectra were measured while monitoring the $Yb^{3+}$ *f-f* emission ($\lambda_{em}$ = 980 nm) using an integrating sphere to capture all emitted photons.

The NC PLE data in Figures S10 and S11 show that the $Yb^{3+}$ *f-f* luminescence is sensitized by the $CsPbCl_3$ host, and that the PLQY remains essentially constant as the excitation wavelength is scanned across the $CsPbCl_3$ absorption spectrum. Furthermore, no new sub-bandgap features are observed in the PLE spectra, even at the highest NC concentration.

Similar to the scenario with high-optical-density NCs, bulk-crystal PLE spectra with direct $CsPbCl_3$ photoexcitation are strongly influenced by being in the high-optical-density limit. To illustrate, Figure S12 plots PLE spectra collected for the 2% $Yb^{3+}$:$CsPbCl_3$ SC sample of the main text in three different powder forms. For the first PLE spectrum, small but macroscopic crystal fragments that were generated during sample cleavage were gathered and used for the PLE measurement. This sample gave a strongly red-shifted first PLE maximum, and a heavily distorted PLE response at shorter excitation wavelengths. This sample was then gently ground twice to reduce crystallite size, and the PLE spectrum was remeasured after each grinding. Reducing the crystallite size dramatically changed the PLE spectrum, reducing and then eliminating these artifacts. In the finest powder, the PLE spectrum now closely resembles the NC PLE spectrum, confirming sensitization of $Yb^{3+}$ by $CsPbCl_3$ in both forms of $Yb^{3+}$:$CsPbCl_3$.

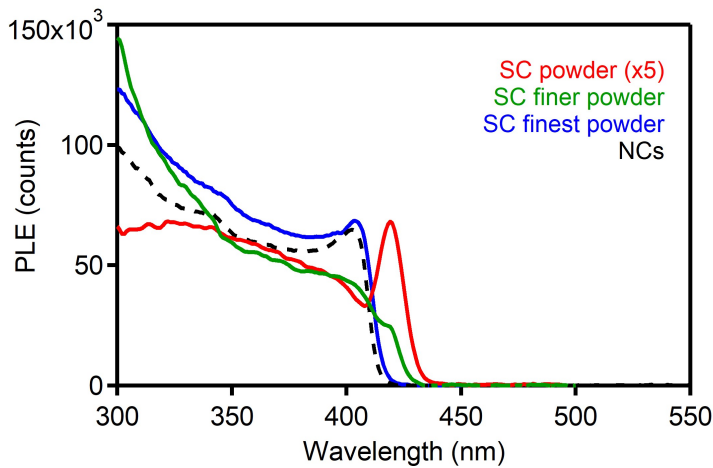

FIG S12. PLE spectra of various forms of the 2% $Yb^{3+}$:$CsPbCl_3$ SC sample described in the main text, measured without an integrating sphere. The PLE spectrum resembles the NC PLE spectrum more and more closely as the grain size decreases, and PLE artifacts are essentially eliminated in the finest powder form. ($\lambda_{em}$ = 985 nm)